 \documentclass[12pt]{iopart}

\usepackage{iopams}

\usepackage{graphicx}
\usepackage{dcolumn}
\usepackage{bm}

\usepackage{color}
\usepackage{ulem}
\begin{document}

\title[Coulomb-dipole transition in mesoscopic electron-hole bilayers]{On the Coulomb-dipole transition in mesoscopic classical and quantum electron-hole bilayers}

\author{P Ludwig$^1$, K Balzer$^1$, A Filinov$^1$, H Stolz$^2$ and M Bonitz$^1$}

\address{$^1$ Institut f\"ur Theoretische Physik und Astrophysik, \\
Christian-Albrechts-Universit\"{a}t zu Kiel, D-24098 Kiel, Germany}

\address{$^2$ Institut f\"ur Physik, Universit\"at Rostock,
D-18051 Rostock, Germany}

\ead{ludwig@theo-physik.uni-kiel.de}

\begin{abstract}
We study the Coulomb-to-dipole transition which occurs when the separation $d$ of an electron-hole bilayer system is varied with respect to the characteristic in-layer distances. An analysis of the classical ground state configurations for harmonically confined clusters with $N\leq30$ reveals that the energetically most favorable state can differ from that of two-dimensional pure dipole or Coulomb systems. Performing a normal mode analysis for the  $N=19$ cluster it is found that the lowest mode frequencies exhibit drastic changes when $d$ is varied. Furthermore, we present quantum-mechanical ground states for $N=6$, $10$ and $12$
spin-polarized electrons and holes. We compute the single-particle energies and orbitals in
self-consistent Hartree-Fock approximation over a broad range of layer separations and coupling strengths between the limits of the ideal Fermi gas and the Wigner crystal.
\end{abstract}

\maketitle

\renewcommand{\vec}[1]{\mathbf{#1}}

\section{Introduction}
Self-organized structure formation, in particular Coulomb crystallization \cite{bonitz08}, is among the most exciting cooperative phenomena in the field of charged many-particle systems.
In the case of finite, parabolically confined systems extensive experimental and theoretical work on various types of two- and three-dimensional systems has revealed that in the strong coupling limit charged particles can arrange themselves in a highly ordered crystalline state with a nested shell structure. Examples are ions in Paul and Penning traps \cite{WinelandEtal,DubinEtal}, dusty plasmas \cite{astra99}-\cite{suga07} and electrons in quantum dots and wells \cite{bedanov-coulomb}-\cite{landman07}. For these also called artificial atoms Mendeleev-type periodic tables were found including characteristic occupation numbers, shell closures and unusually stable ”magic” configurations. For a recent overview see \cite{bonitz08}.
Recently there is growing interest in two-dimensional (2D) \textit{dipolar} macroscopic systems \cite{palo02}-\cite{kalman07}
as well as finite size dipolar (quantum) clusters in small-scale confinement potentials \cite{belousov-dipole}-\cite{zoller07}. While in particular the ground state and dynamical properties of 2D mesoscopic pure Coulomb and pure dipole interacting particle ensembles in parabolic confinement potentials
are well understood, the behaviour of \textit{real three-dimensional} electron-hole double layer systems, where the dipole approximation is not valid, is still poorly investigated. This despite the fact that the additional degree of freedom, i.e. the layer separation $d$, is expected to allow for a variety of interesting new effects which are due to the possibility of tuning the effective in-layer interaction potential.

The results presented in this paper are applicable to semiconductor heterostructures and coupled quantum dots as well as to molecular systems, where the dipole moment of the charge carriers and thus the interaction strength is tunable, e.g. \cite{zoller07,Xtrap}.\footnote{Another natural source of confinement arises in low-dimensional semiconductor structures from defects and well width fluctuations. This leads to local potential minima for the charge carriers causing localization of free and bound charges (excitons, biexcitons, trions), e.g. \cite{afilinov04a,afilinov04b,bracker05}.}
For a consistent formulation we concentrate on the problem of two vertically coupled symmetric layers containing parabolically confined, spin-polarized electrons and holes of identical particle number $N_e=N_h=N$ and effective masses $m^*_e=m^*_h=m^*$, respectively. The underlying Hamiltonian is
\begin{equation} \label{ham0}
\hat H=\hat H_e+\hat H_h-\hat H_{e-h} \;,
\end{equation}
with the intra- and interlayer contributions
\begin{eqnarray}\label{ham1}
\hat H_{e(h)} = \sum\limits_{i=1}^{N_{e(h)}} \left( -\frac{\hbar^2}{2 m^*_{e(h)}} \nabla_{{\vec r}_i}^2 + \frac{m^*_{e(h)}}{2}\omega_{0}^2 {\vec r}_i^2+ \sum\limits_{j=i+1}^{N_e(h)} \frac{e^2}{4 \pi \varepsilon \sqrt{({\vec r}_i- {\vec r}_j)^2}}  \right) \;,\\ 
\hat H_{e-h}= \sum\limits_{i=1}^{N_e}\sum\limits_{j=1}^{N_h} \frac{e^2}{4 \pi \varepsilon \sqrt{({\vec r}_i- {\vec r}_j)^2+d^2}} \;,
\label{ham2}
\end{eqnarray}
where the electrons (e) and holes (h) are confined to planes
of zero thickness which are a distance $d$ apart. The 2D vectors
${\vec r}_{i(j)}$ are the in-plane projections of the particle coordinates, $e$ the elementary charge and $\varepsilon$ the static permittivity. The  strength of the confinement is controllable by the trap frequency $\omega_{0}$.

The most fascinating property of this system is that the effective in-layer particle interaction changes with the interlayer separation $d\,$: from Coulomb interaction at large $d$, where both layers are decoupled, to dipole interaction at small $d\rightarrow0$, where the attractive interlayer interaction leads at low temperature to vertical electron-hole coupling and formation of vertically aligned dipoles --- excitons. On the other hand, at intermediate values of $d$, when the repulsive intra- and attractive interlayer interaction energies according to equations (\ref{ham1}) and (\ref{ham2}) are comparable, the system shows a real three-dimensional behaviour.
In reference \cite{ludwig03} it was reported that, as a consequence of the Coulomb-dipole transition, the considered system can exhibit structural changes of its ground state shell configuration when $d$ is varied.

In section~\ref{groundstates} we extend these results and present a systematic study of the classical ground states, varying $d$ for mesoscopic clusters with $N\leq30$ particles in each layer. Further, we extract the fundamental dynamical features in the case of weak excitation by solving the dynamical (Hessian) matrix for the ground state configurations found in section~\ref{groundstates}. Doing this, in section~\ref{modes} we discuss the $d$-dependence of the collective $N$-particle modes for the $N=19$ cluster. Here we highlight the close relationship between structural and collective dynamical cluster properties as rotation of shells and vortices.
In section \ref{quantum} we extend the analysis to fermionic e-h quantum bilayers utilizing a self-consistent Hartree-Fock ansatz. In particular, Coulomb-to-dipole transition induced (critical) quantum phenomena are presented for the clusters with $N=6, 10$ and $12$ electrons and holes. The results include the $N$-particle densities and the single-particle spectrum and orbitals as function of coupling strength $\lambda$ and layer separation~$d$.

\section{Classical ground state transitions} \label{groundstates}
The classical ground state corresponding to the equations (\ref{ham0}) to (\ref{ham2}) is described by the Hamiltonian $H=H_e+H_h-H_{e-h}$ without the kinetic energy, i.e. 
\begin{eqnarray}\label{ham_cl}
H_{e(h)}= \sum\limits_{i=1}^{N}{\vec r}_i^2 + \sum\limits_{i<j}^{N} \frac{1}{\sqrt{({\vec r}_i- {\vec r}_j)^2}} \; , \; H_{e-h}= \sum\limits_{i=1}^{N_e} \sum\limits_{j=1}^{N_h} \frac{1}{\sqrt{({\vec r}_i- {\vec r}_j)^2+d^2}} \;.
\end{eqnarray}
This dimensionless form is obtained applying the transformation rules $\{r \rightarrow r/r_0$, $E \rightarrow E/E_0$, $d \rightarrow d/r_0\}$ with the characteristic length $r_0 = (e^2/2 \pi \varepsilon m \omega_0^2)^{1/3}$ and energy $E_0 = (m \omega_0^2 e^4/32 \pi^2 \varepsilon^2)^{1/3}$. Note that model (\ref{ham_cl}) contains no explicit dependence on the trap frequency $\omega_0$. The considered classical model system in the ground state is completely defined by only two parameters: the particle number $N$ and the layer separation $d$, which also influences the in-layer density.

The ground state configuration is the energetically lowest of all possible \textit{stable} states, whose number rapidly increases with $N$, and all these have to be found and checked.
This task is complicated, since many of the different stable states are energetically close, requiring high-accuracy computations.
A systematic search for the global minimum-energy structure in the
$4N$-dimensional configuration space was performed by means of an optimized molecular dynamics annealing technique utilizing an adaptive step size control  \cite{ludwig05,ludwig03}.
For each value of $N$ and $d$ the annealing process was repeated for a large ($N$- and $d$-dependent) number of times. At this a slow (long) annealing process ensures to find the lowest-energy state with high probability. The critical points of structural transitions $d_{cr}$ were identified as crossing points of the energies of the lowest-energy states as function of layer separation $d$.

\begin{table}[bt]
\begin{center}
\begin{tabular}{c |c|c c c|c}
\hline
\hline
$N$      & Coulomb              & Bilayer                                        & $d_{cr}$  &$E_{cr}/N$  & Dipole   \\
\hline
  5      &  5                   & no transition                                    &           &           &   5      \\
  6      & (1,5)                & no transition                                    &           &           &  (1,5)   \\
  7      & (1,6)                & no transition                                    &           &           &  (1,6)   \\
  8      & (1,7)                & no transition                                    &           &           &  (1,7)   \\
  9      & (2,7)                & no transition                                    &           &           &  (2,7)   \\
  10     & (2,8)                & \underline{(2,8)} $\rightarrow$ (3,7)            &  $1.0116$ &   3.9167  &  (3,7)   \\
  11     & (3,8)                & no transition                                    &           &           &  (3,8)   \\
  12     & \underline{(3,9)}    & \underline{(3,9)} $\rightarrow$ (4,8)            &  0.9528   &   4.3463  &  \underline{(3,9)}   \\
         &                      & (4,8) $\rightarrow$ \underline{(3,9)}           &  0.3253   &   2.1293  &          \\
  13     & (4,9)                & no transition                                    &           &           &  (4,9)   \\
  14     & (4,10)               & no transition                                    &           &           &  (4,10)  \\
  15     & (5,10)               & no transition                                    &           &           &  (5,10)  \\
  16     &\underline{(1,5,10)}  & no transition                                    &           &           & \underline{(1,5,10)} \\
  17     &(1,6,10)              & no transition                                    &           &           & (1,6,10) \\
  18     &(1,6,11)              & no transition                                    &           &           & (1,6,11) \\
  19     &\underline{(1,6,12)}  & \underline{(1,6,12)} $\rightarrow$ (1,7,11)     & 2.182     & 9.1882    & \underline{(1,6,12)} \\
         &                      & (1,7,11) $\rightarrow$ \underline{(1,6,12)}     & 0.417     & 3.5697    &          \\
  20     &(1,7,12)              & no transition                                    &           &           & (1,7,12) \\
  21     &(1,7,13)              & (1,7,13) $\rightarrow$ (2,7,12)                 & 3.429     & 11.6283   & (2,7,12) \\
  22     &(2,8,12)              & no transition                                    &           &           & (2,8,12) \\
  23     &(2,8,13)              & (2,8,13) $\rightarrow$ (3,8,12)                 & 2.436     & 10.9959   & (3,8,12) \\
  24     &(3,8,13)              & no transition                                    &           &           & (3,8,13) \\
  25     &(3,9,13)              & no transition                                    &           &           & (3,9,13) \\
  26     &(3,9,14)              & (3,9,14) $\rightarrow$ (4,9,13)                 & 2.173     & 11.4266   & (4,9,13) \\
  27     &(4,9,14)              & no transition                                    &           &           & (4,9,14) \\
  28     &(4,10,14)             & no transition                                    &           &           & (4,10,14)\\
  29     &(4,10,15)             & (4,10,15) $\rightarrow$ (5,10,14)               & 2.142     & 12.2357   & (5,10,14)\\
  30     &\underline{(5,10,15)} & \hspace{0.5pc} \underline{(5,10,15)} $\rightarrow$ (1,5,10,14) & 0.616     & 6.3934    & \underline{(5,10,15)}\\
         &                      & \hspace{-0.7pc}(1,5,10,14) $\rightarrow$ \underline{(5,10,15)} & 0.243     & 3.3410    &          \\
\hline
\hline
\end{tabular}
\end{center}
\caption{Ground state shell structures for 2D Coulomb, bilayer and dipole clusters of $N$ particles in a parabolic confinement.
The arrows indicate the direction of the ground state transition from large to small values of $d$. Magic (commensurate) shell configurations are underlined. For $N\leq5$ only a single shell is populated for all values of $d$.
For all configurational  transitions the critical layer separation $d_{cr}$ as well as the corresponding total energy per composite dipole $E_{cr}/N$ is given. Note that the binding energy $1/d$ which ensures the exact vertical alignment of the electron-hole pairs is excluded from the energy values as it is independent of the cluster configuration.

}\label{gstable}
\end{table}

Extending the analysis of \cite{ludwig03} we obtained a periodic table for the particle numbers $N\leq30$ including all structural transitions occurring when $d$ is changed, see table \ref{gstable}. In the limits of pure dipole and Coulomb interaction our results are in full agreement with those of reference \cite{belousov-dipole} and  references \cite{bedanov-coulomb, kongPRE}, respectively.\footnote{In reference \cite{bedanov-coulomb} the  ground state for $N=29$ was erroneously given as (5,10,14). This was corrected in reference \cite{kongPRE}.}
Analyzing the clusters $N\leq18$ only transitions for $N=10$ and $N=12$, reported in reference \cite{ludwig03}, are found. Due to the much larger configurational space, and thus accordingly higher number of low-energy metastable states, for the clusters $N=19\ldots30$ in total $6$ particle numbers reveal Coulomb-dipole transitions: $N=19$, $21$, $23$, $26$, $29$ and $30$. In particular, two transition types are identified:
\begin{enumerate}
\item[(A)]  While for the majority of the investigated clusters the ground state shell configuration of the single layer Coulomb and dipole case are identical, for $N=10$, $21$, $23$, $26$ and $29$ this is not the case. When changing from a long-range Coulomb to a short-range dipole interaction a higher particle number on the inner shell becomes favourable. A similar trend is also known from 2D \cite{LaiPRE99,kongJPhys} and 3D \cite{bonitzPRL06,henning06} Yukawa-clusters when the screening strength is increased. \footnote{The effect is due to the radial balance of total internal $F^{int}$ and external $F^{ext}$ forces on each particle. In contrast to Coulomb, short-range (dipole or Yukawa) forces  do contribute  to $F^{ext}$ which requires a higher density towards the center to stabilize the cluster matching $F^{int}=F^{ext}$. For details see \cite{henning06}.}
\item[(B)] A second type of transition is found for $N=12$, $19$ and $30$ that cannot be concluded from different shell occupations in both limits of $d\,$: At large values of $d$ again a transition of type (A) takes place, which increases (decreases) the particle number on the inner (outer) shell when $d$ is reduced.
But interestingly, at small values of $d$ a second kind of transition to a sixfold-coordinated, commensurate particle configuration is found allowing for an energetically more favourable \textit{closed packing} of the composite dipoles. Such symmetry-induced re-entrant configuration changes are only observed in cases where highly symmetric, ``magic'' configurations with a bulk-like triangular structure are involved.
\end{enumerate}
These findings coincide with those for single layer statically screened Coulomb systems. Here a change from the long-range Coulomb towards a short-range Yukawa potential by variation of the screening length leads to analogue ground state transitions for the particle numbers   $N=10$, $12$, $19$ and $N=21$, $23$, $26$, $29$ as reported in reference \cite{kongJPhys} and \cite{LaiPRE99}, respectively.
Further, a comparison of the ground and metastable states of the single layer Coulomb system  (cf. table 1 in reference~\cite{kongPRE} for $N\leq30$) shows that if and only if an energetically close metastable configuration with higher center particle number than the ground exists, in fact, a transition of type (A) in the corresponding bilayer system is found. This underlines the Coulomb-to-dipole transition induced density change effecting configurational transitions of type (A).
In contrast, transitions of type (B) are geometry induced supporting an equally distant, closed packed particle arrangement.

Among all transitions, the most interesting are those of type (B). As an example, we study the $N=19$ cluster. Here between $d=0.417$ and $d=2.182$, the ``magic'' configuration (1,6,12) is replaced by the configuration (1,7,11) which possesses a much lower orientational order \cite{schweigert95}.
Therefore, it is interesting to analyze the normal modes of this cluster and their dependence on $d$.

\section{Collective $N$-particle modes} \label{modes}
Starting from the ground state configurations given in section \ref{groundstates}, we are interested in the collective excitation behaviour in dependence on $d$. Here we will focus on the cluster with $N=19$ where, upon changing $d$, finite size effects are expected to play a key role as the ground state structure changes between the hexagonally ordered (1,6,12) configuration  and the (1,7,11) circular ring structure as discussed in section \ref{groundstates}.

To derive the dynamical properties in the limit of weak excitations we perform a normal mode analysis \cite{fil03,schweigert95,elliott,balzer06}.
For small particle displacements $\vec{u}(t) = \vec{r}(t) - \vec{R}$ around their ground state position $\vec{R}$, expansion of the potential energy $U$, equation (\ref{ham0}), around $\vec{R}$ leads to
\begin{eqnarray} \label{Uentw}
U(\vec{r})= U_0 + \sum_i^{2N} \underbrace{\left. \frac{\partial U}{\partial r_i} \right|_{\vec{R}} }_{=\,0} u_i +  \frac{1}{2} \sum_{i,j}^{2N}\underbrace{\left. \frac{\partial^2 U}{\partial r_i \partial r_j} \right|_{\vec{R}} }_{=: \,{\cal H}_{ij}} u_i u_j + \ldots \quad,
\end{eqnarray}
where $U_0$ is the minimum potential energy. In the stationary states the linear (force) term vanishes and the second-order partial derivatives provide the  elements ${\cal H}_{ij}$ of the $2\times2N$ Hessian matrix.
In the frame of the harmonic approximation the resulting cluster dynamics is given as a superposition of these collective (normal) modes statistically weighted according to the eigenvalues of ${\cal H}$ which are proportional to the squared mode oscillation frequencies $\omega_i^2$. In the following these eigenfrequencies will be given in units of $\omega_0/\sqrt{2}$.

\begin{figure}[t]
\begin{center}
\includegraphics[width=1.0\textwidth]{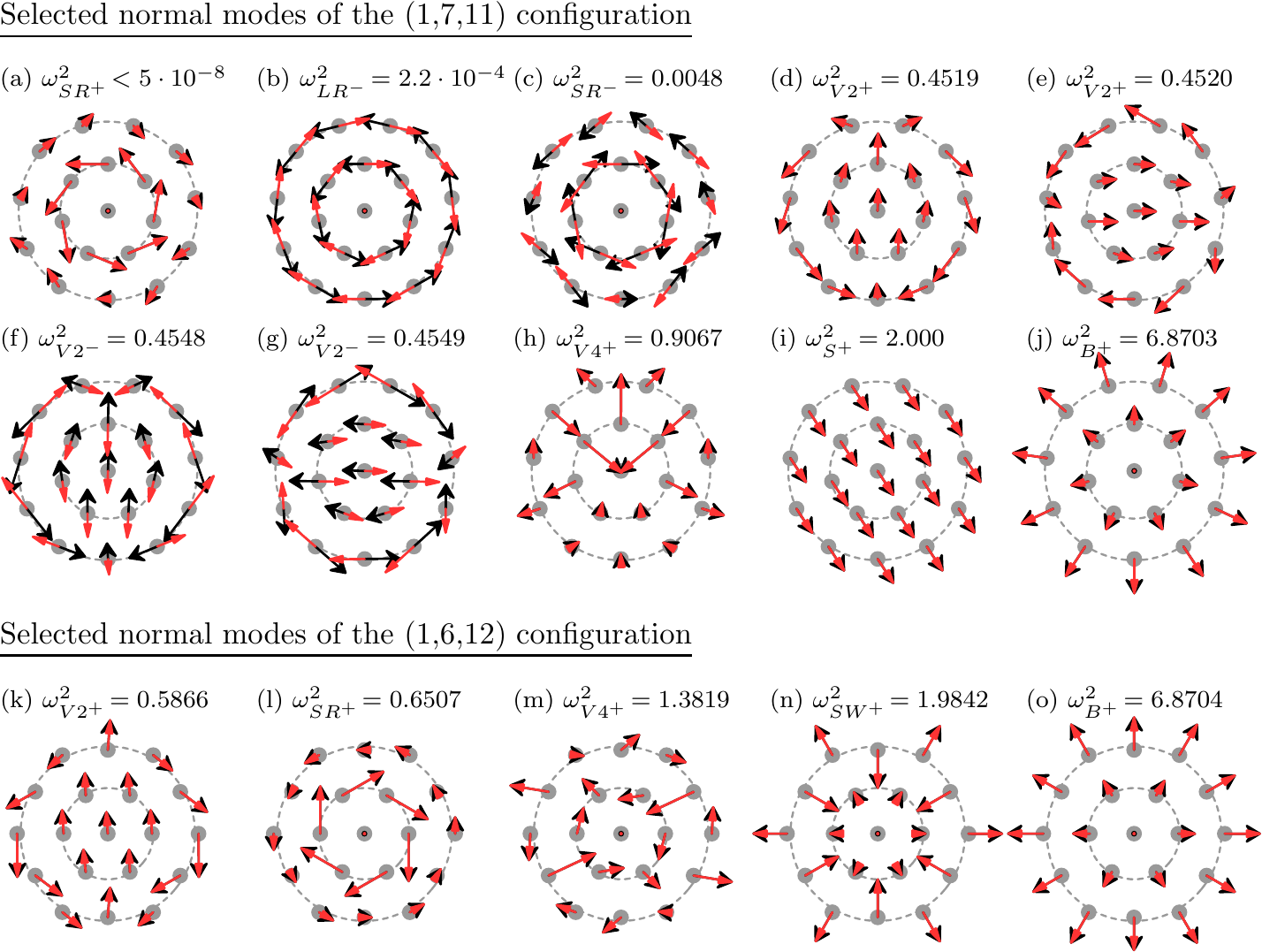}
\caption{Top view of the eigenvectors of selected characteristic and  low-energetic normal modes for the $N=19$ cluster at $d=d_{cr}=2.182$
(ordered by frequency, cf. numbers above the figures). The points mark the particle positions. The different shaped (and colored) arrow heads are assigned to the normal mode eigenvectors in the two different layers and indicate direction and amplitude of particle motion. Modes with in-/anti-phase motion of 2 layers are labeled with a $^+/ ^-$ sign, respectively . \\
\underline{Top rows:} Eigenvectors of the (1,7,11) configuration:
a) inter-shell rotation (SR$^+$),
b) anti-phase layer rotation (LR$^-$),
c) anti-phase inter-shell rotation (SR$^-$),
d) and e) in-phase vortex pairs (V2$^+$),
f) and g) anti-phase vortex pairs (V2$^-$),
h) asymmetric in-phase 4-vortex mode ($V4^+$),
i) sloshing mode ($S^+$),
j) breathing mode ($B^+$). \\
\underline{Bottom row:} Eigenvectors of the (1,6,12) configuration:
k) in-phase vortex pair (V2$^+$),
l) in-phase inter-shell rotation (SR$^+$),
m) in-phase 4-vortex mode ($V4^+$),
n) in-phase transverse surface wave ($SW^+$),
o) breathing mode ($B^+$).
}\label{modeplot}
\end{center}
\end{figure}

\subsection{Classification of normal modes}\label{class}
As a result of the eigenmode computation we obtain for each stable configuration of the $N=19$ cluster a complete set of $76$ eigenvalues and eigenvectors. A selection of characteristic and energetically low-lying eigenvectors for $d\!=\!d_{cr}\!=\!2.182$, i.e. intermediate between Coulomb and dipole regime, is given in figure \ref{modeplot}.
As shown in reference \cite{fil03}, in dipolar bilayer systems the total number of modes can be divided in two types which will be distinguished by the following nomenclature:
\begin{enumerate}
\item[($^+$)] labels modes with \textit{in-phase} collective particle motion in both layers, see figure~\ref{modeplot}~a), d), e), h) to o), and
\item[($^-$)] labels modes with \textit{anti-phase} motion of both layers, see figure~\ref{modeplot}~b), c), f) and g).
\end{enumerate}

Consider first the top rows of figure~\ref{modeplot} which show the eigenvectors of the normal modes of the (1,7,11) configuration. The energetically lowest collective particle motion is in all cases the center of mass cluster rotation mode --- the \textit{in-phase} layer rotation LR$^+$. The eigenfrequency of this directed rotation is  $\omega=0$ as for this motion there is no restoring force. Beside this (trivial) mode there are three additional rotational modes:   \textbf{a)} inner versus outer inter-shell rotation SR$^+$, \textbf{b)} the \textit{anti-phase} rotation of both layers LR$^-$ and \textbf{c)} anti-phase inter-shell rotation SR$^-$.

Another set of low frequency modes are four vortex pair modes: \textbf{d)}  \textit{in-phase} vortex pair V2$^+$ and \textbf{e)} perpendicular oriented vortex pair V2$^+$, \textbf{f)} and \textbf{g)} two anti-phase vortex pairs.
In the present, isotropically confined 2D system rotationally \textit{asymmetric} modes are typically two-fold degenerate with respect to the spatial alignment of the vectors, cf.  d), e) and f), g), respectively. This leads to the fact that, taking into account the two possible phasings of relative particle motion in both layers, a majority of mode types occures as a set of four. Considering this, in the following only one mode per set of four is shown as for the rotational asymmetric, low-energy mode \textbf{h)} which has the interesting feature that it supports a single particle exchange between the inner and outer shell, i.e. a transition from the (1,7,11) to the (1,6,12) configuration.

In the case of pure radial eigenvectors, such as the (in-phase) breathing mode \textbf{j)} as coherent radial motion (compression/expansion) of all particles, there exists one pair of modes only. In addition to j) there is an anti-phase breathing mode B$^-$ with frequency $\omega^2_{B^-}=7.9522$. Further ``universal modes`` that are independent of particle number and configuration is the  center of mass sloshing mode \textbf{i)} with trap frequency $\omega_0$ which has a corresponding anti-phase shear or dipole oscillation mode S$^-$ (each two-fold degenerate).

For all these modes a corresponding mode of the (1,6,12) configuration is found. In particular: \textbf{k)} the V2$^+$-mode, \textbf{l)} the mode of inter-shell rotation SR$^+$, \textbf{m)} an energetically low V4$^+$-mode, here supporting a center directed transition of a particle on the outer shell, and further two examples of radial modes, \textbf{n)} a transverse surface wave and \textbf{o)} the breathing mode.

\subsection{Change of normal mode spectrum with layer separation}

\begin{figure}[!h]
\begin{center}
\includegraphics[width=20.50pc]{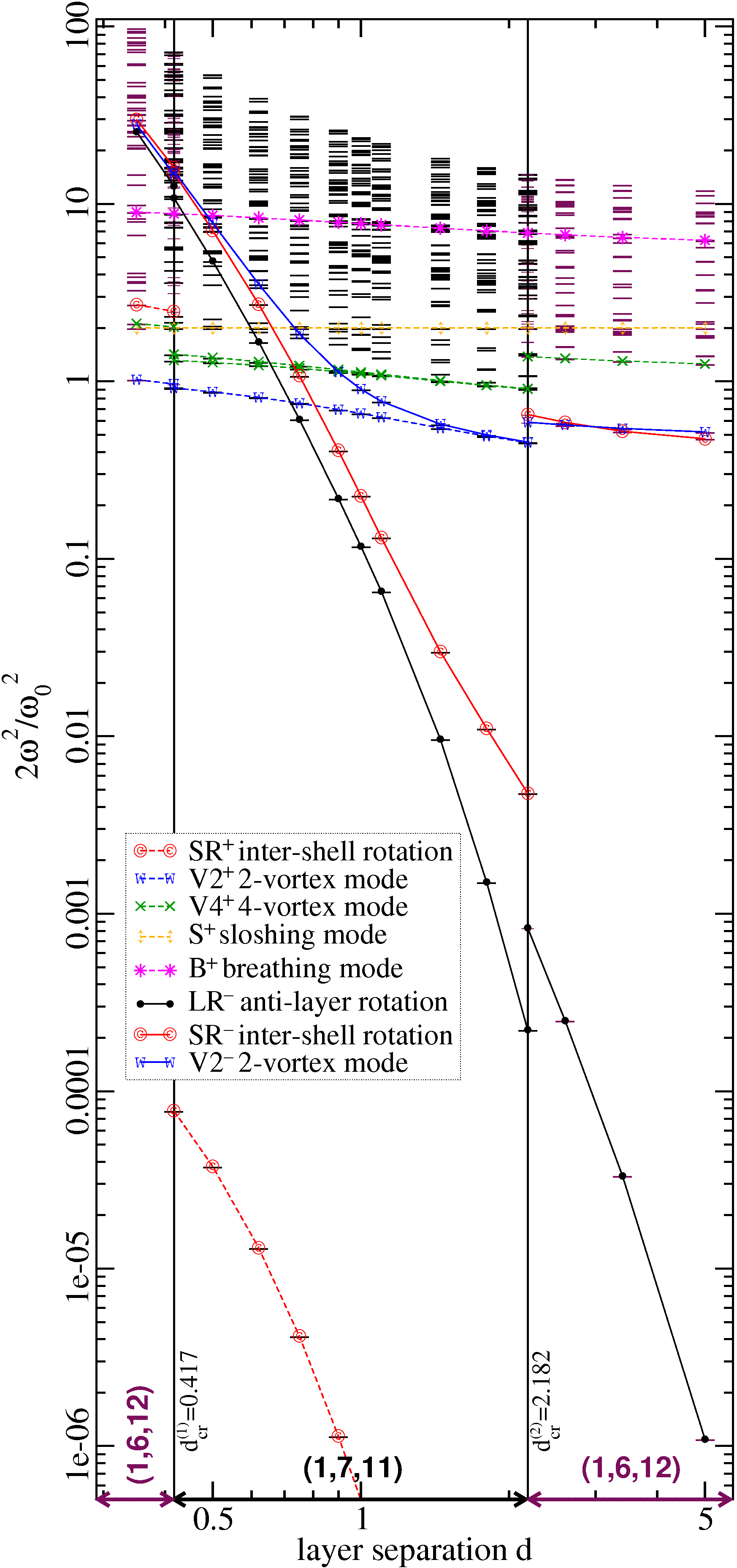}
\caption{Complete normal mode spectrum for $N=19$ as function of layer separation $d$. At $d^{(1)}_{cr}=0.417$ the ground state configuration changes from $(1,6,12)$ to $(1,7,11)$ and at $d^{(2)}_{cr}=2.182$ from $(1,7,11)$ to $(1,6,12)$ resulting in a qualitative change of the mode frequencies. The eigenvectors of the selected modes are visualized in figure~\ref{modeplot}. Modes with in-phase (anti-phase) oscillation of both layers are plotted with dashed (solid) lines. Note that the SR$^+$ mode continues in the range $1<d<2.182$ with a value smaller than $5\cdot10^{-7}$. For notation of modes see figure \ref{modeplot}.}
\label{spect}
\end{center}
 \end{figure}

After the classification of the collective modes we now consider the oscillation frequency dependence on the layer separation $d$ of the $N=19$ cluster, see figure \ref{spect}. Of special interest are thereby the two configuration changes of the ground state and their effect on the collective dynamical cluster properties.

Starting at small values of $d$, an increase of the e-h separation leads to a growing cluster size due to a stronger in-layer particle repulsion resulting from a change of the effective interaction from dipole to Coulomb. This implicates a gradual decrease of the mode eigenfrequencies with $d$ since the coupling of all $2N$ particles becomes less rigid and the restoring forces weaken. Only the two-fold degenerate center of  mass oscillations are found to be constant at $\omega^2_{S^+}=2$, independent of the interlayer coupling strength or even configuration changes. Confirming reference \cite{fil03}, the breathing frequency gradually  proceeds from $\omega_{B^+}^2=10$ in the limit of dipoles ($d \rightarrow 0$), to a value of  $\omega_{B^+}^2=6$ in the limit of decoupled layers ($d \rightarrow \infty$). Moreover, modes supporting a transition from the (1,6,12) to the (1,7,11) state and vice versa, i.e. the eigenmodes h) and m) in figure~\ref{modeplot}, are  found at low frequencies, i.e., at low excitation energies.

As discussed in section \ref{groundstates}, the ground state transitions for the $N=19$ cluster occur at the critical values of $d^{(1)}_{cr}=0.417$ and  $d^{(2)}_{cr}=2.182$ and are accompanied by abrupt spectrum transformations.
The strongest effect is observed for the in-phase inter-shell rotation SR$^+$  with a remarkable jump of the mode frequency $\omega^2_{SR^+}$ by more than four orders of magnitude.
This decrease can be explained by comparing the SR$^+$ mode eigenvectors of the (1,7,11) and (1,6,12) configurations, see figure \ref{modeplot} a) and l). In the latter case, the oscillation vectors of all particles on the inner shell are directed towards particle positions on the outer shell
which strongly increases the restoring forces in the case of the (1,6,12) configuration resulting in a much higher frequency $\omega_{SR^+}^{(1,6,12)}$ than $\omega_{SR^+}^{(1,7,11)}$.
The exceptional low frequency $\omega_{SR^+}^{(1,7,11)}$ agrees with results for single layer Coulomb crystals. In reference \cite{schweigert95} the minimal (non-zero) excitation frequency for (1,7,11) and the comparable, non-magic (1,7,12) configuration was reported to be that of the inter-shell rotation with $\omega_{SR}^{2}\approx10^{-8}$. 
Confirming this, quantum Monte Carlo simulations \cite{afilinov01} revealed that the orientational inter-shell melting temperature of the incommensurate (1,7,12) configuration is much lower than for the high symmetric (1,6,12) structure. In particular, a 9 (!) orders of magnitude difference of the orientational melting temperatures and critical densities of both configurations was found. This shows that the given classical results are of practical relevance also for quantum systems at moderate densities. 

Moreover, with respect to the dipole-to-Coulomb transition we found that in the dipole regime at small $d$ the corresponding modes with in-phase and anti-phase oscillation of both layers are energetically clearly separated, cf.  SR$^\pm$  and V2$^\pm$ in figure \ref{spect}. Energetically lowest are the two (degenerate) in-phase vortex pair oscillations V2$^+$. With a gradual transition to the limit of uncoupled layers, the e-h attraction and thus the oscillation frequencies of the anti-phase modes are strongly reduced and converge towards the values of the corresponding in-phase modes. This is found for the V2$^-$ and V2$^+$ modes around $d=2$ and for the SR$^-$ and SR$^+$ modes for $d>2.182$. As a consequence of the layer decoupling, the LR$^-$ anti-phase layer rotation becomes the energetically lowest of the anti-phase modes. This indicates that the primary mechanism of decoupling of the electron and hole layers is the interlayer rotation LR$^-$.

\section{Ground states and single-particle spectrum of quantum bilayers}\label{quantum}

In this section we present an extension of the classical results of section \ref{groundstates} to quantum bilayers.
Here, in contrast to the classical simulations, the ground state kinetic energy does not vanish even in the limit of temperatures $T\rightarrow0$ resulting in a finite spatial extension of the particle orbitals on the scale of the whole $N$-particle cluster. Hence, fermionic quantum features such as exchange effects (Pauli exclusion principle) must be included.

In order to treat the e-h bilayer system of equations (\ref{ham0}-\ref{ham2}) quantum mechanically, we introduce the dimensionless coupling parameter $\lambda$ of a harmonically confined quantum system which relates the characteristic Coulomb energy
$E_C=e^2/(4\pi\epsilon x_0)$ to the characteristic confinement energy $E^*_0=\hbar\omega_0$
\begin{eqnarray}
\lambda=\frac{E_C}{E^*_0}=\frac{e^2}{4\pi\epsilon x_0 \hbar\omega_0}=\frac{x_0}{a_B}\;,
\end{eqnarray}
where $x_0=\sqrt{\hbar/(m\omega_0)}$ denotes the oscillator length and $a_B=4 \pi \epsilon \hbar^2 /(m e^2)$ is the effective electron (hole) Bohr radius. Thus Hamiltonian (\ref{ham_cl}) including the kinetic energy can be rewritten in dimensionless form
\begin{eqnarray}\label{ham2a}
\hat H_{e(h)} &=& \frac{1}{2} \sum\limits_{i=1}^{N}(-\nabla^2_i + {\vec r}_i^2) +  \sum\limits_{i<j}^{N} \frac{\lambda}{\sqrt{({\vec r}_i- {\vec r}_j)^2}} \; ,   \\
\label{ham2b}
\hat H_{e-h} &=& \sum\limits_{i=1}^{N_e} \sum\limits_{j=1}^{N_h}
 \frac{\lambda}{\sqrt{({\vec r}_i- {\vec r}_j)^2+d^{*2}}}\;,
\end{eqnarray}
using the transformation $\{r \rightarrow r/x_0$, $E \rightarrow E/E^*_0$, $d^* \rightarrow d/x_0\}$.
 Note that $r$ and $d^*$ are measured in units of $x_0$ and thus explicitly depend on the confinement frequency $\omega_0$. The  characteristic energies and length scales of the classical (section \ref{groundstates}) and quantum system are related by
\begin{eqnarray} \label{conversion}
\frac{E_0}{E^*_0} = (\lambda^2/2)^{1/3} \; , \quad \frac{r_0}{x_0} = (2\lambda)^{1/3} \;,
\end{eqnarray}
so that the layer separations used in the Hamiltonians~(\ref{ham2}) and (\ref{ham2b}), respectively, are related by $d^*=(2\lambda)^{1/3} \, d$.

In the limit $\lambda\rightarrow0$, both, electrons and holes behave as an ideal trapped Fermi gas independent of the layer separation $d^*$. For $\lambda\rightarrow\infty$, it is $x_0/a_B\gg 1$, and quantum effects vanish. Thus one recovers classical behavior and shell configuration changes which coincide with those in table \ref{gstable}. At finite $\lambda$, however, intra- and interlayer interactions, together with the parabolic confinement, give rise to a complex quantum many-body problem, which is subject of the following investigation. In the considered quantum case, ground state properties depend on the two parameters $d^*$ and $\lambda$. Therefore, the question arises if the additional degree of freedom will induce additional structural changes. To answer this question, we performed self-consistent Hartree-Fock (SCHF) calculations of two coupled electron and hole layers of zero thickness, which are discussed in the next two subsections.

\subsection{Second quantization formulation}

\newcommand{\mean}[1]{\langle{#1}\rangle}

In order to derive mean-field type equations for the e-h bilayer, we rewrite the exact Hamiltonian (\ref{ham2a},\ref{ham2b}) in the second-quantized form $\hat{H}=\hat{H}_e+\hat{H}_h-\hat{H}_{e-h}$, where
\begin{eqnarray}
\label{ham2ndq_a}
 \hat{H}_{e(h)}&=&\int\!\textup{d}^2r\, \hat{\psi}_{e(h)}^\dagger(\vec{r})\,h_0(\vec{r})\,\hat{\psi}_{e(h)}(\vec{r})\\
&&+\frac{1}{2}\int\!\!\!\int\textup{d}^2r\textup{d}^2\bar{r}\,\hat{\psi}_{e(h)}^\dagger(\vec{r})\,\hat{\psi}_{e(h)}^\dagger(\bar{\vec{r}})\,\frac{\lambda}{\sqrt{(\vec{r}-\bar{\vec{r}})^2}}\,\hat{\psi}_{e(h)}(\bar{\vec{r}})\,\hat{\psi}_{e(h)}(\vec{r})\;,\nonumber\\
\label{ham2ndq_b}
 \hat{H}_{e-h}&=&\int\!\!\!\int\textup{d}^2r\textup{d}^2\bar{r}\,\hat{\psi}_{e}^\dagger(\vec{r})\,\hat{\psi}_{h}^\dagger(\bar{\vec{r}})\,\frac{\lambda}{\sqrt{(\vec{r}-\bar{\vec{r}})^2+d^{*2}}}\,\hat{\psi}_{h}(\bar{\vec{r}})\,\hat{\psi}_{e}(\vec{r})\;,
\end{eqnarray}
with $h_0(\vec{r})=\frac{1}{2}(-\nabla^2+\vec{r}^2)$ denoting the single-particle energy. Further, $\hat{\psi}^{(\dagger)}_{e(h)}(\vec{r})$ is the annihilation (creation) operator of spin-polarized electrons and holes at space point $\vec{r}$ which satisfy the fermionic anti-commutation relations ${[\hat{\psi}_{e(h)}(\vec{r}),\hat{\psi}_{e(h)}^\dagger(\bar{\vec{r}})]}_{+}=\delta(\vec{r}-\bar{\vec{r}})$ and ${[\hat{\psi}^{(\dagger)}_{e(h)}(\vec{r}),\hat{\psi}_{e(h)}^{(\dagger)}(\bar{\vec{r}})]}_{+}=0$ where $[\hat{A},\hat{B}]_{+}=\hat{A} \hat{B}+\hat{B} \hat{A}$. In a Hartree-Fock approach \cite{Echen07}, the four field operator products entering equations (\ref{ham2ndq_a}) and (\ref{ham2ndq_b}) are approximated by sums over double products $\hat{\psi}^\dagger_{e(h)}\hat{\psi}_{e(h)}$ weighted by the generalized electron (hole) density matrix $\rho_{e(h)}(\vec{r},\bar{\vec{r}})=\mean{\hat{\psi}_{e(h)}^\dagger(\vec{r})\hat{\psi}_{e(h)}(\bar{\vec{r}})}_{e(h)}$, where the expectation value (ensemble average) is defined as $\mean{\hat A}_{e(h)}=\textup{Tr}\,{\hat{\rho}_{e(h)}\hat A}$. More precisely, with $\eta,\xi\in\{e,h\}$, the 4-operator products are approximated as
\begin{eqnarray}
\label{HFapprox}
 &&\,\,\hat{\psi}^\dagger_{\eta}(\vec{r})\,\hat{\psi}^\dagger_{\xi}(\bar{\vec{r}})\,\hat{\psi}_{\xi}(\bar{\vec{r}})\,\hat{\psi}_{\eta}(\vec{r})\\
&\approx&\,+\,{\rho}_{\eta}(\vec{r},\vec{r})\,\hat{\psi}_{\xi}^\dagger(\bar{\vec{r}})\,\hat{\psi}_{\xi}(\bar{\vec{r}})\,+\,{\rho}_{\xi}(\bar{\vec{r}},\bar{\vec{r}})\,\hat{\psi}_{\eta}^\dagger(\vec{r})\,\hat{\psi}_{\eta}(\vec{r})\nonumber\\
&&\,-\,\delta_{\eta\xi}\left[{\rho}_{\eta}(\vec{r},\bar{\vec{r}})\,\hat{\psi}_{\xi}^\dagger(\bar{\vec{r}})\,\hat{\psi}_{\xi}(\vec{r})\,+\,{\rho}_{\xi}(\bar{\vec{r}},\vec{r})\,\hat{\psi}_{\eta}^\dagger(\vec{r})\,\hat{\psi}_{\eta}(\bar{\vec{r}})\right]\;.\nonumber
\end{eqnarray}
Here, the first two terms constitute the Hartree term whereas the last two denote the Fock (exchange) contribution. The Kronecker delta $\delta_{\eta\xi}$ assures that there is no exchange between electrons and holes which is
due to the different physical nature of electrons and holes (different energy bands).
Inserting the approximate expression (\ref{HFapprox}) into (\ref{ham2ndq_a},\ref{ham2ndq_b}) allows for an effective one-particle description according to
\begin{eqnarray} \label{HFeff1}
 \hat{H}_{e(h)}&=&\!\!\int\!\!\!\int\textup{d}^2r\textup{d}^2\bar{r}\,\hat{\psi}_{e(h)}^\dagger(\vec{r})\left\{h_0(\vec{r})\,\delta(\vec{r}-\bar{\vec{r}})+\Sigma^{\mathrm{HF}}_{e(h)}(\vec{r},\bar{\vec{r}})\right\}\hat{\psi}_{e(h)}(\bar{\vec{r}})\;,\\ \label{HFeff2}
\hat{H}_{e-h}&=&\!\!\int\!\!\!\int\textup{d}^2r\textup{d}^2\bar{r}\,\hat{\psi}_{e}^\dagger(\vec{r})\left\{\Sigma^{\mathrm{HF}}_{e-h}(\vec{r},\bar{\vec{r}})+\Sigma^{\mathrm{HF}}_{h-e}(\vec{r},\bar{\vec{r}})\right\}\,\hat{\psi}_{h}(\bar{\vec{r}})\;,
\end{eqnarray}
with the Hartree-Fock (HF) self-energies
\begin{eqnarray}
\Sigma^{\mathrm{HF}}_{e(h)}(\vec{r},\bar{\vec{r}})=\lambda\int\textup{d}^2r'\,\frac{{\rho}_{e(h)}(\vec{r}',\vec{r}')}{\sqrt{(\vec{r}'-\vec{r})^2}}\,\delta(\vec{r}-\bar{\vec{r}})\,-\,\lambda\,\frac{{\rho}_{e(h)}(\vec{r},\bar{\vec{r}})}{\sqrt{(\vec{r}-\bar{\vec{r}})^2}}\;,\\
\Sigma^{\mathrm{HF}}_{e-h(h-e)}(\vec{r},\bar{\vec{r}})=\lambda\int\textup{d}^2r'\,\frac{{\rho}_{h(e)}(\vec{r}',\vec{r}')}{\sqrt{(\vec{r}'-\vec{r})^2+d^{*2}}}\,\delta(\vec{r}-\bar{\vec{r}})\;.
\end{eqnarray}
For computational reasons it is convenient to introduce a basis representation for the electron (hole) field operators,
\begin{eqnarray}
\label{basisexpansion}
\hat{\psi}^{(\dagger)}_{e(h)}(\vec{r})&=&\sum_{i}\varphi^{(*)}_i(\vec{r})\,\hat{a}^{(\dagger)}_{e(h),i}\;,\hspace{1.5pc}i\in\{0,1,2,\ldots\}\;,
\end{eqnarray}
where the one-particle orbitals or wave functions $\varphi_i(\vec{r})$ form an orthonormal complete set and $\hat{a}^{(\dagger)}_{e(h),i}$ is the annihilation (creation) operator of a particle on the level $i$. Applying the basis expansion (\ref{basisexpansion}) to the equations (\ref{HFeff1}) and (\ref{HFeff2}) leads to the matrix representation of the bilayer Hamiltonian (\ref{ham0}) which will be given in the following section, cf. equations (\ref{HFham}-\ref{HFham3}).

\subsection{Self-consistent Hartree-Fock simulation technique} \label{SCHFtech}
In matrix representation, the mean-field Hamiltonian for the bilayer system corresponding to the initial equations (\ref{ham0}-\ref{ham2}) reads
\begin{eqnarray}\label{HFham}
h_{ij}^{e(h)}&=&h_{ij}^{0}\,+\,h_{ij}^{e-e(h-h)}\,-\,h_{ij}^{e-h(h-e)}\;,\\
\label{HFham2}
h_{ij}^{e-e(h-h)}&=&\lambda\sum_{kl}\left(w^{e-e(h-h)}_{ij,kl}\,-\,w^{e-e(h-h)}_{il,kj}\right)\rho^{e(h)}_{kl}\;, \hspace{0.5pc}\\
\label{HFham3}
h_{ij}^{e-h(h-e)}&=&\lambda\sum_{kl}w^{e-h(h-e)}_{ij,kl}\,\rho^{h(e)}_{kl}\;,
\end{eqnarray}
with the single-particle (orbital) quantum numbers $i$ and $j$ ($k$ and $l$), $h_{ij}^{e(h)}$ being the electron (hole) total energy, $h_{ij}^{0}$ the single-particle (kinetic and confinement) energy and $h_{ij}^{e-e(h-h)}$ ($h_{ij}^{e-h(h-e)}$) the intra (inter) layer interactions in mean-field approximation.
 Further,
$\rho_{ij}^{e(h)}=\mean{\hat a _{e(h),i}^\dagger\,\hat a _{e(h),j}}$
denotes the zero-temperature density matrix of electrons and holes
with respect to the one-particle basis $\varphi_i(\vec{r})$.
In equation~(\ref{HFham2}) both the  Hartree and the Fock contribution appear, whereas in equation~(\ref{HFham3}) only the Hartree term enters.

The explicit expression for the single-electron (-hole) integral is
\begin{eqnarray}
\label{hij0}
h_{ij}^{0}&=&\frac{1}{2}\int \textup{d}r^2\,\varphi^*_i(\vec{r})(-\nabla^2+\vec{r}^2)\varphi_j(\vec{r})\;,
\end{eqnarray}
and the two-electron (two-hole) and electron-hole integrals are given by
\begin{eqnarray}
\label{wijklee} \label{wijklee}
w_{ij,kl}^{e-e(h-h)}&=&\!\int\!\!\!\int \textup{d}^2r d^2\bar{r}\, \frac{\varphi^*_i(\vec{r})\,\varphi^*_k(\bar{\vec{r}})\,\varphi_j(\vec{r})\,\varphi_l(\bar{\vec{r}})}{\sqrt{(\vec{r}-\bar{\vec{r}})^2+\alpha^{*2}}}\;,\\
\label{wijkleh}
w_{ij,kl}^{e-h(h-e)}&=&\!\int\!\!\!\int \textup{d}^2r d^2\bar{r}\, \frac{\varphi^*_i(\vec{r})\,\varphi^*_k(\bar{\vec{r}})\,\varphi_j(\vec{r})\,\varphi_l(\bar{\vec{r}})}{\sqrt{(\vec{r}-\bar{\vec{r}})^2+d^{*2}}}\;,
\end{eqnarray}
where $\alpha^*\rightarrow0$ is utilized to avoid the Coulomb singularity for $\vec{r}\rightarrow\bar{\vec{r}}$.
A small parameter of $\alpha^*\lesssim0.01$ has been found to show convergence for all quantities of interest. Details will be given elsewhere \cite{balzer08}.

For numerical implementation of the SCHF procedure yielding the eigenfunctions~$\phi_i^{e(h)}(\vec{r})$ (Hartree-Fock orbitals) and eigenenergies~$\epsilon_i^{e(h)}$ (Hartree-Fock energies) of Hamiltonian~(\ref{HFham}), we have chosen the orthonormal Cartesian (2D) harmonic oscillator states
\begin{eqnarray} \label{ostates}
\varphi_{m,n}(\vec{r})&=&\frac{e^{-(x^2+y^2)/2}}{\sqrt{2^{m+n}\,m!\,n!\,\pi}}\,\,{\cal H}_m(x)\,{\cal H}_n(y)\;,
\end{eqnarray}
with single-particle quantum numbers~$i=(m,n)$, $\vec{r}=(x,y)$, the Hermite polynomials ${\cal H}_m(x)$ and $(m+1)$-fold degenerate energy eigenvalues~$\epsilon_{m,n}=m+n+1$ where $m,n\in\{0,1,2,\ldots\}$. The Hartree-Fock orbitals, expanded in the form
\begin{eqnarray}
\phi_i^{e(h)}(\vec{r})=\sum_{j=0}^{n_b-1} c^{e(h)}_{ji}\,\varphi_j(\vec{r})\;,
\end{eqnarray}
with coefficients $c^{e(h)}_{ij}\in\mathbb{R}$ and respective energies $\epsilon_i^{e(h)}$, are obtained by iteratively solving the self-consistent Roothaan-Hall equations \cite{roothaanhall}
\begin{eqnarray}\label{Roothaan-Hall}
\sum_{k=0}^{n_b-1} h_{ik}^{e(h)}\,c^{e(h)}_{kj}\,-\,\epsilon^{e(h)}_j\,c^{e(h)}_{ij}&=&0\;,
\end{eqnarray}
at fixed dimension $n_b\times n_b$ ($i=0,1,\ldots n_b-1$)  according to standard techniques, for details see e.g. \cite{Echen07} and references therein. The resulting electron (hole) density $\rho^{e(h)}_{d^*,\lambda}(\vec{r})$ corresponding  to given values of $d^*$ and $\lambda$ is defined as
\begin{eqnarray}
\rho^{e(h)}_{d^*,\lambda}(\vec{r})=\sum_{k=0}^{N-1}\phi_{k,d^*,\lambda}^{e(h)}(\vec{r})=\sum_{k=0}^{N-1}\sum_{l=0}^{n_b}c^{e(h)}_{lk,d^*,\lambda}\,\varphi_l(\vec{r})\;,
\end{eqnarray}
where each Hartree-Fock orbital~$k$ is occupied by a single particle only.

For the e-h bilayers with  $N\leq10$, we used $n_b=50$ of the energetically lowest oscillator functions $\varphi_{m,n}(\vec{r})$ to expand the Hartree-Fock orbitals, for $N=12$ we took $n_b=55$ which was sufficient to obtain convergent results.
Due to the electron-hole attraction the cluster size is reduced compared to that of a single layer Coulomb cluster. This favours the use of a moderate number of basis functions to ensure convergence. \footnote{Note that the additional center particle in the case of $N=19$ strongly increases the cluster size so that essentially more basis functions ($n_b\gtrapprox90$) are required to ensure convergence. As the problem determining the two-particle integrals, equations (\ref{wijklee}) and (\ref{wijkleh}), scales with ${\cal O}( n_b^4)$ a computation is limited by memory requirements.}

\subsection{Transition from the ideal Fermi gas towards the classical limit} \label{clsec}

\begin{figure}[t]
\begin{center}
\includegraphics[width=0.95\textwidth]{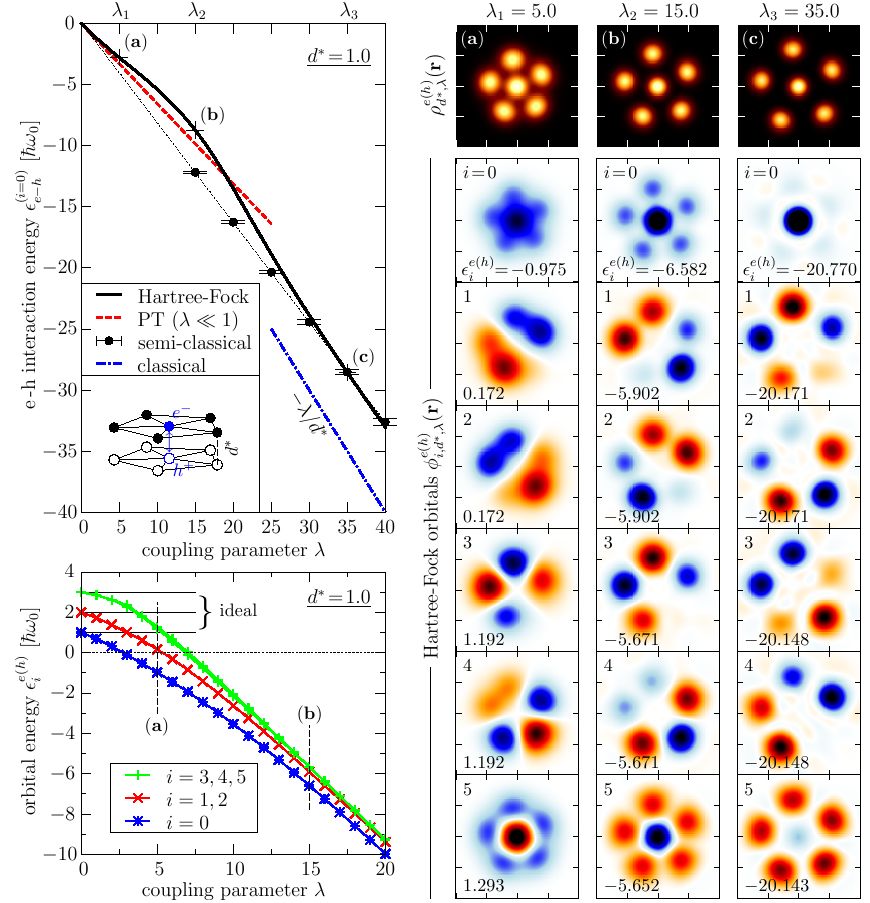}
\caption{Ground state of the $N=6$ cluster as function of interaction strength $\lambda$ for fixed layer separation $d^*=1.0$.
\underline{Right:} Accumulated $N$-particle density $\rho_{d^*,\lambda}^{e(h)}(\vec r)$, on top, and corresponding single-particle HF orbitals $\phi_{i,d^*,\lambda}^{e(h)}(\vec{r})$ for three different coupling parameters $\lambda$. The different signs of the wave function (blue and orange) are separated by white areas of zero amplitude, whereas areas of maximum amplitude are black. Note that the six high-density spots of the $N$-particle density do not necessarily correspond to the single particles themselves as the configuration appears as a superposition of all orbitals.
\underline{Left:} (top) Electron-hole interaction energy $\epsilon_ {e-h}^{(i=0)}$, equation (\ref{e-h-binding}), of the center electron and hole states for different approximations and (bottom) HF energy of occupied levels $\epsilon^{e(h)}_i$ as function of $\lambda$.
}\label{fig.N6}
\end{center}
\end{figure}

Aim of this part is to investigate the transition from a strongly degenerate quantum system, i.e.  $\lambda=0$, to the classical limit $\lambda \rightarrow \infty$. To give a reasonable estimate for the range at which the classical ground state results become valid, we consider a system with $N=6$ electrons and holes at an intermediate layer separation of $d^*=1.0$. Of special interest will be the central spot of the (1,5) configuration which can most directly be assigned to a classical particle position.

In contrast to the classical results the Hartree-Fock (HF) calculations fully take into account the wave nature of electrons and holes. The quantum many-body effects are evident already at $\lambda=0$. In the classical case, the total energy in the ground state is zero (all particles sit in the bottom of the trap). In the quantum case, this is prevented by the Pauli principle. Orbital-resolved HF calculations  as function of coupling parameter $\lambda$ are displayed in figure~\ref{fig.N6}.
Here, the right panel shows the $N$-particle density $\rho_{d^*,\lambda}^{e(h)}(\vec r)$ and the six populated single-particle orbitals $\phi^{e(h)}_{i,d^*,\lambda}(\vec r)$ for moderate ($\lambda_1=5.0$), intermediate ($\lambda_2=15.0$) and strong ($\lambda_3=35.0$) coupling. The SCHF results reveal that, in particular for small values of $\lambda$, obviously several orbitals contribute collectively to the different high-density spots which unambiguously determine the cluster configuration.

Concerning the lowest orbital $i=0$, with an increase of $\lambda$, the overlap with the higher orbitals vanishes and the wave function becomes localized when $\lambda$ exceeds a value of $35$. In contrast, in the investigated range of $\lambda\leq40$ the other particles remain, independently of the observed density modulation, delocalized as can be seen on the orbital pictures.
The transition towards the limit of strong correlations can be estimated from the e-h-interaction energy
\begin{eqnarray}\label{e-h-binding}
\epsilon_ {e-h}^{(i=0)}(\lambda)=-\!\int\!\!\!\int\textup{d}^2r\textup{d}^2\bar{r}\,
|\phi^e_{i=0}(\vec r)|^2\, \frac{\lambda}{\sqrt{(\vec{r}-\bar{\vec{r}})^2+d^{*2}}} \, |\phi^h_{i=0}(\bar\vec r)|^2 \;,
\end{eqnarray}
of the electron and hole in the lowest orbital. The upper diagram in the left panel of figure~\ref{fig.N6} displays the $\lambda$-dependence for four different approximations.
For the ideal system, $\lambda=0$, electron and hole are not bound and $\epsilon_ {e-h}^{(i=0)}$ vanishes. The black solid line shows the interaction energy (\ref{e-h-binding}) obtained from the SCHF simulations which for $\lambda\ll1$ agrees with perturbation theory (PT) where a linear $\lambda$-dependence follows from substituting the ideal wave function $\varphi_{0,0}(\vec r)$, see equation (\ref{ostates}), for $\phi^{e(h)}_{i=0}(\vec r)$ in equation (\ref{e-h-binding}).

For $\lambda\gg1$ a semi-classical result can be derived. Starting from the classical ground state configuration (1,5) the outer particles, together with the confinement, create an effective potential for both center particles which can be harmonically approximated. The direct quantum mechanical solution of the harmonic problem provides a finite Gaussian electron (hole) extension of width $\sigma=\sigma_e=\sigma_h$. Hence, the e-h-interaction energy (\ref{e-h-binding}) of the inner particles can be computed in a semi-classical way using $\phi^{e(h)}_{i=0}(\vec{r})=\left(\sigma/\pi\right)^{1/4}e^{-\sigma(x^2+y^2)/2}$. In the strongly correlated regime, starting at  $\lambda\geq30$, the semi-classical and SCHF solution coincide very well.

However, in an intermediate coupling range, $\lambda\approx15$, the e-h interaction energy is reduced compared to the semi-classical solution which reflects the fact that the orbital $i=0$ substantially deviates from a Gaussian, cf. the five side maxima of the orbital $i=0$ for $\lambda_2=15$ in figure~\ref{fig.N6}. With increase of $\lambda$ this Gaussian becomes more and more peaked describing the transition to the classical limit $|\phi_{i=0}^{e(h)}(\vec r)|^2\rightarrow\delta(\vec r)$.\footnote{In the mean-field Hamiltonian (\ref{HFeff1},\ref{HFeff2}) the classical limit is obtained by replacing $\rho_{e(h)}(\vec r, \vec r')\rightarrow\delta(\vec r-\vec r')\sum_{i=1}^{N_{e(h)}}\delta(\vec r-\vec r_i)\;$. } Despite the good agreement with the semi-classical approximation, in the whole investigated range of $\lambda<40$ the system is found to be essentially non-classical. This becomes evident by comparing with the pure classical result $\epsilon_{e-h}^{(i=0)}=-\lambda/d^*$ which neglects any finite particle extension.
Concerning all populated HF orbitals the transition towards the classical limit with increasing $\lambda$ is shown in the lower left diagram of figure~\ref{fig.N6} in terms of the orbital energies $\epsilon_{i=(m,n)}^{e(h)}$. As mentioned in section~\ref{SCHFtech} the harmonically confined ideal Fermi gas ($\lambda=0$) is $(m+1)$-fold degenerate with $m,n\in\{0,1,\ldots\}$. Around $\lambda\geq15$ the energy of the outer particles converges towards a five-fold degenerate energy which is separated from the (lower) energy of the center particle.

\subsection{Quantum ground state configurations and structural transitions for $N=10$} \label{secN10}
Beside the higher numerical effort of a single SCHF computation compared to its classical analogue, a complete study of the ground states requires, in addition to $d$ and $N$, the exploration of $\lambda$ as a third degree of freedom.
To overcome this problem and to reduce the task, we limit our investigation to the two-shell clusters $N=10$ and $N=12$ which were found to exhibit rich ground state properties in the classical limit.

The analysis was done by systematically scanning the phase diagram for fixed values of $d^*$ ranging from $0.1\ldots10.0$. For each of these $d^*$ values we start from the ideal system at $\lambda=0$ and increment the coupling parameter stepwise by $\delta\lambda=0.05$. The convergence of each step is ensured by an adaptive, precision controlled iteration number with up to 2500 iterations of the Roothaan-Hall equations (\ref{Roothaan-Hall}) per increment $\delta\lambda$. The described procedure allows for a systematic investigation of the phase diagram by a gradual transition from the ideal Fermi to the strongly coupled system.
To verify the results obtained, the ground states with respect to individual points in the phase diagram were recomputed by starting from a random distribution as well as by decreasing the temperature of an initial (high temperature) thermal distribution~\cite{balzer08}. All procedures are found to yield the same HF orbitals (energies) and thus the same $N$-particle densities and shell structures.

\begin{figure}[t]
\includegraphics[width=\textwidth]{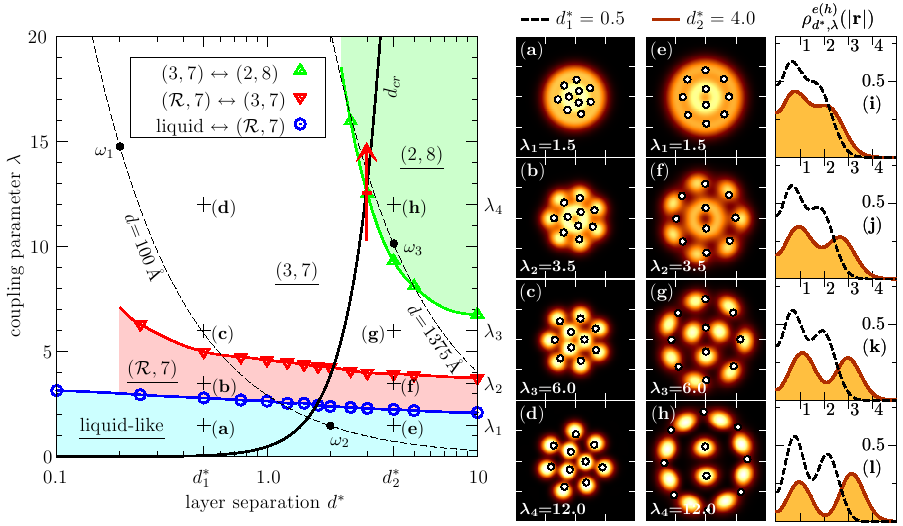}
\caption{ \underline{Left:} $(\lambda,d^*)$-phase diagram for the $N=10$ bilayer in HF approximation. The configuration $({\cal R},7)$ means delocalization of charges on the inner ring $\cal R$. The black solid line indicates the classical ground state transition $(3,7)\rightarrow(2,8)$ which occurs  at $d_{cr}=1.0116 r_0$ from left to right. The red arrow points out an inverse transition compared to the classical $(2,8)\rightarrow(3,7)$ crossing. The two dashed lines indicate the path when changing $\omega_0$ at fixed layer separation $d$ for a germanium ($\varepsilon=16\varepsilon_0$, $m^*_{e(h)}=0.25 m_e$) quantum-well structure, see equation~(\ref{lambda_d*}); $\omega_1=926$ GHz, $\omega_2=9.26$ THz, $\omega_3=98$ GHz.\\
\underline{Right:}
Electron (hole) density $\rho^{e(h)}_{d^*,\lambda}(\vec{r})$ at characteristic points marked (a) to (h) in the phase diagram. The side length of the contour plots is $9 x_0$. The open circles mark the corresponding classical ground state particle  positions. The rightmost column displays the corresponding angle-averaged radial density profiles for $d^*=0.5$ ($4.0$), dashed (red) line.}
\label{fig.N10}
\end{figure}

The results for the $N=10$ cluster are presented in figure \ref{fig.N10}. The ground state phase diagram can be divided into four domains (left panel of figure \ref{fig.N10}):

\begin{enumerate}
\item At small $\lambda$ a weakly correlated degenerate Fermi liquid is observed within each layer (blue area in the left figure). The observed electron (hole) density is rotationally symmetric and exhibits non-monotonic radial modulations of an (nearly) ideal trapped Fermi gas. The proper density distributions for $d_1^*=0.5$ and $d_2^*=4.0$ are shown in figures (a) and (e), respectively.

\item At higher $\lambda$ two shells separate, see points (b), (f) within the red area in the phase diagram and the corresponding density profile (j). While on the inner ring the  electron (hole) density is still isotropic, the density on the outer shell becomes angle-modulated and reveals seven high-density spots. The integrated position probability density on the inner and outer shells is close to 3 and 7, respectively. The configuration will be referred to in the following as $({\cal R},7)$ as on the inner ring ${\cal R}$ no localized density peaks in $\rho^{e(h)}_{d^*,\lambda}(\vec r)$ are present. \textit{Hence the nomenclature does not indicate the particle numbers, but the number of distinct density peaks}, as the particle orbitals are delocalized over the entire cluster, see discussion in section~\ref{clsec} and \ref{secorbitals}.

\item Further increase of the coupling parameter leads to more pronounced (concentric) shells. In particular, the inner radial density decreases which is accompanied by the formation of angular density modulation, see figures (c) and (g). The shell configuration is found to be (3,7).

\item  At a certain value $\lambda_{cr}(d^*)$, the bilayer system jumps from the $(3,7)$ into the  $(2,8)$ shell configuration (green area in the phase diagram), see figures (g) $\rightarrow$ (h).
\end{enumerate}

The general behaviour of (i) - (iii) is independent of the layer separation $d^*$.
The localized (3,7) configuration (iii) emerges in two steps by rotational symmetry breaking from the Fermi liquid (i) maintaining a higher density on the inner than on the outer ring.
However, an increase of $d^*$ beyond unity leads, by weakening of the interlayer attraction, to a repulsive intralayer and thus Coulomb-dominated coupling.
Consequently, the cluster size increases, compare the density plots of figure~\ref{fig.N10}~(a) vs. (e), (b) vs. (f), etc.$\;$.
Moreover, for a fixed $\lambda\gg1$, the dipole-to-Coulomb transition towards the strongly correlated Coulomb regime induces the $(2,8)$ shell configuration \cite{ludwig03} which is observed when  $d^*$ is increased from $0.5$ to $4.0$, see figure (d) vs. (h). This transition reduces the inner-shell density, see figure~\ref{fig.N10} right (red vs. dashed lines).

Further, at a fixed $d^*>2.0$ an increase of $\lambda$ leads to a purely coupling induced configuration change (3,7) $\rightarrow$ (2,8), see details in section~\ref{secorbitals}.
For $d^*=10$, both layers are already weakly coupled and become completely decoupled when $d^*$ is further increased. Consequently, the critical (blue, red and green) curves in the phase diagram converge towards horizontal lines.
Note, that $d^*$ is measured in units of $x_0$ and thus depends on the confinement frequency $\omega_0$. This implies for an experimental setup, e.g. a double quantum-well heterostructure with fixed physical layer separation $d$, that one traces hyperbolas of the form
\begin{eqnarray}
\label{lambda_d*}
\lambda(d^*)=\frac{d\,e^2 m^*_{e(h)}}{4\pi\varepsilon\,\hbar^2}\,\frac{1}{d^*(\omega_0)}\;,
\end{eqnarray}
when changing the trap frequency $\omega_0$, see the dashed lines in the phase diagram of figure~\ref{fig.N10}. The larger the physical layer separation $d$ (or effective particle mass $m^*_{e(h)}$), the more the hyperbola shifts to larger values of $d^*$.
Interestingly, e.g., for a germanium based quantum well, at fixed layer separation $d=1375$\AA, the ground state structure of the quantum bilayer can be externally controlled by change of $\omega_0$ only.

A comparison of the classical particle positions (open circles in figures (a)-(h)), according to equations (\ref{conversion}), with the shells and high-density spots of the HF calculations plotted in figure~\ref{fig.N10} reveals a good agreement. Larger cluster sizes compared to the classical case for small $\lambda$ are explained by repulsive fermionic exchange interactions. Further, the bold black line in the phase diagram indicates the classical transition from (3,7) to (2,8) which occurs at $d_{cr}=1.0116 r_0$ when crossing the line from left to right. It is found that the classical line gives a reasonable estimate also for the transition in the quantum bilayer system.
Hence the trend, found in section \ref{groundstates} for the classical bilayer system, of center density reduction with increasing $d$ also holds in the case of a strongly correlated quantum system, where the orbitals extend over several classical particle positions.
In the classical limit, i.e.~at very large $\lambda$ (outside of figure~\ref{fig.N10}), the configuration boundary $(3,7)\leftrightarrow (2,8)$ (green curve) and the classical result (black curve) converge.
Nevertheless, for intermediate values of $\lambda$ the red arrow indicates a remarkable  point in the phase diagram where the structural transition in the classical and quantum bilayer proceeds in opposite direction. The single-particle orbitals for this transition will be analyzed in section \ref{secorbitals}.

Further, an unusual (2,8) configuration is shown for $\lambda_4=12.0$ and $d^*_2=4.0$ in figure~\ref{fig.N10}(h), where the particle arrangement differs from the classical system. Such a configuration was also found in reference \cite{ludwig03} for a classical single layer system with $1/r^\alpha$ pair interaction and $\alpha\leq0.94$. Thus, the anomalous configuration underlines the effect of the Fermi repulsion in addition to the intralayer Coulomb interaction. However, an increase of $\lambda$ leads to a reduction of the Fermi effect and wave function overlap and a (2,8) configuration corresponding to the classical one is found.

\subsection{Quantum ground state configurations and structural transitions for $N=12$}

\begin{figure}[t]
\begin{center}
\includegraphics[width=0.85\textwidth]{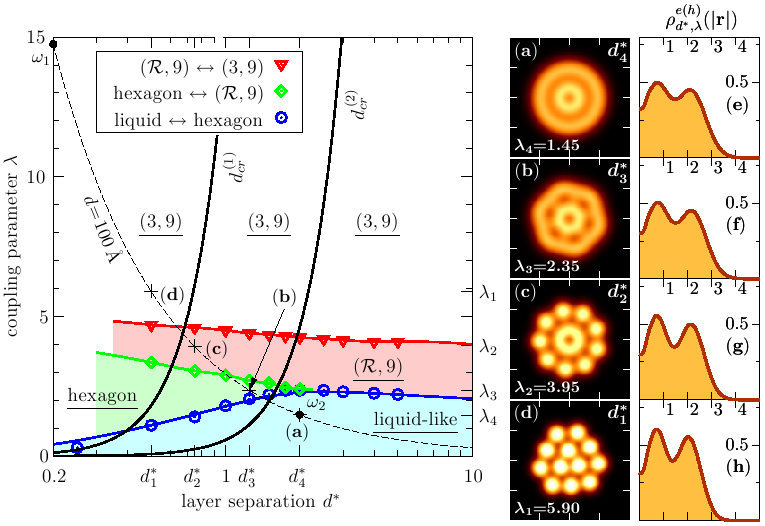}
\caption{$(\lambda,d^*)$-phase diagram showing the quantum shell structures found
for the $N=12$ bilayer in HF approximation. The shown electron (hole)
densities $\rho^{e(h)}_{d^*,\lambda}(\vec{r})$ corresponding to tuples (d)
 $(d^*_1,\lambda_1)=(0.5,5.9)$, (c) $(d^*_2,\lambda_2)=(0.75,3.95)$, (b) $(d^*_3,\lambda_3)=(1.25,2.35)$ and (a) $(d^*_4,\lambda_4)=(2.0,1.45)$. The frequencies $w_{1,2}$ are as indicated in figure \ref{fig.N10}. The two black solid lines indicate the classical configuration transitions $(3,9)\rightarrow(4,8)$ and $(4,8)\rightarrow(3,9)$ at $d_{cr}^{(1)}=0.9528r_0$ and $d_{cr}^{(2)}=0.3253r_0$, respectively, from left to right. In the investigated range $\lambda\leq15$ these transitions were not observed in the quantum bilayer.
The right two columns show the (radial) density of the four points (a) to (d) marked in the phase diagram for $d=100$\AA.}\label{fig.N12}
\end{center}
\end{figure}

In figure \ref{fig.N12} we present the $(\lambda,d^*)$-phase diagram for $N=12$ electrons and holes analogous to figure \ref{fig.N10} for $N=10$. At fixed (physical) layer separation $d=100$\AA~one passes through four different domains of the phase diagram when the trap frequency is decreased from $\omega_2$ towards $\omega_1$ (see left panel of figure \ref{fig.N12}):
\begin{enumerate}
\item Analogously to the $N=10$ cluster at small $\lambda$, a weakly correlated circular symmetric Fermi liquid exists within each layer, see point (a) in the blue area of the phase diagram.
\item A decrease of the trap frequency to point (b) is accompanied by a structural change to a 6-fold rotational cluster symmetry with an outer shape exhibiting hexagonal symmetry. This phase only establishes in the regime of a short-range in-layer potential, i.e.  $d^*\leq2$. In the Coulomb case of weakly coupled layers this liquid-like state is not found.
\item If the confinement strength is further reduced, see point (c), the cluster passes over to a 9-fold rotational symmetry. While in the cluster core a ring  ${\cal R}$ of delocalized density is observed, the outer nine high-density spots are situated on a perfectly circular ring, which reproduces the symmetry of the external confinement potential.
\item In the limit of small $d^*$ and $\lambda\rightarrow\infty$, see figure (d), where the in-layer interaction becomes extremely short-range, a commensurate closed packed structure with $3-$fold rotational symmetry similar to that known from classical dipole systems \cite{ludwig03} is found.
\end{enumerate}
Consequently, during the coupling-induced transition from (i) to (iv) the cluster size decreases slightly as the effective in-layer interaction becomes short-ranged. In analogy to $N=10$, the liquid-like state (i) as well as the $({\cal R},9)$ configuration (iii) are found for all values of $d^*$. The additional configuration (ii), missing in the case of $N10$, is limited to a range of strong interlayer attraction.

In contrast to  $N=10$, in total two transitions as function of $d$ were found in the classical $N=12$ system, cf. table~\ref{gstable}. However, in the investigated quantum regime, $\lambda\leq15$, we observe no configuration changes corresponding to the classical transitions $(3,9)\leftrightarrow(4,8)$, see black lines $d_{cr}^{(1)}=0.9528r_0$ and $d_{cr}^{(1)}=0.3253r_0$ in figure \ref{fig.N12} left. Hence the two ground state transitions $(3,9)\rightarrow(4,8)$ and $(4,8)\rightarrow(3,9)$ of type (A) and (B), introduced in section \ref{groundstates}, are expected to occur outside of figure \ref{fig.N12} in the (semi-)classical region only.

\subsection{Single-particle orbitals and single-particle spectrum} \label{secorbitals}

\begin{figure*}[t]
\includegraphics[width=\textwidth]{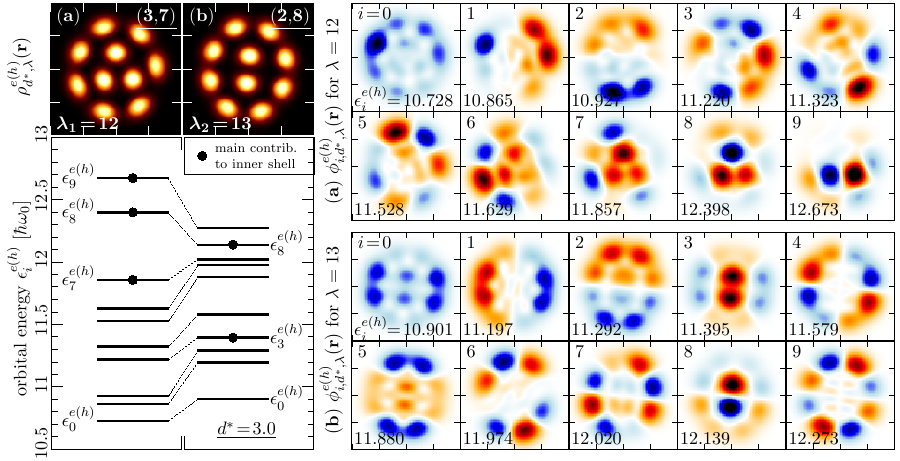}
\caption{\underline{Left:} Hartree-Fock energy eigenvalues $\epsilon_i^{e(h)}$ corresponding to the spatial orbitals $\phi^{e(h)}_{i,d^*,\lambda}(\vec{r})$. (a) Bilayer density $\rho^{e(h)}_{d^*,\lambda}(\vec{r})$ and orbitals for $N=10$ at $d^*=3.0$ and $\lambda_1=12.0$, (b) $\lambda_2=13.0$. The black dots denote the orbitals which contribute most to the inner-shell high-density spots. While the inner shell of the (3,7) configuration is essentially build up from the 3 highest orbitals 7, 8 and 9, the inner shell of the (2,8) configuration is mostly formed from the orbitals 3 and 8.\\
\underline{Right:} Single-particle orbitals $\phi^{e(h)}_{i,d^*,\lambda}(\vec{r})$ for the cases (a) and (b).
The different signs of the wave function (blue and orange) are separated by white areas of zero amplitude, whereas areas of maximum amplitude are black.

}\label{fig.N10spectrum}
\end{figure*}

In both previous subsections we discussed the phase diagram based on the $N$-particle densities. In this part we pursue the question of how the single-particle spectrum evolves during the transition from (3,7) to (2,8) for the $N=10$ cluster, see red arrow in figure~\ref{fig.N10} left. At fixed $d^*=3.0$, the configurational transition occurs when changing the coupling parameter from $\lambda_1=12$ to $\lambda_2=13$. For this transition, the spatially resolved orbitals $\phi_{i,d^*,\lambda}^{e(h)}(\vec{r})$ and the $N$-particle density  $\rho^{e(h)}_{d^*,\lambda}(\vec{r})$ are collected in figure~\ref{fig.N10spectrum} together with the corresponding one-particle HF spectra $\epsilon_i^{e(h)}$ for both coupling parameters $\lambda_{1}$ and $\lambda_{2}$.

As mentioned in section \ref{secN10} the configuration change $(2,8)_{\lambda_1}\leftrightarrow(3,7)_{\lambda_2}$ is reversed along the red arrow in figure~\ref{fig.N10} compared with the respective classical transition.
Similar to the $N=6$ cluster discussed in figure~\ref{fig.N6} the HF orbitals generally extend over several classical particle positions.\\
In situation $(a)$, i.e. $\lambda_1=12$, the energetically highest orbitals $i=7$, $8$ and $9$ contribute most to the inner-shell density showing three high-density spots. On the other hand, in $(b)$, i.e. $\lambda_2=13$, the orbitals are completely rearranged with the two inner-shell density spots being now formed mainly from the orbitals 3 and 8, leading to embedded orbital energies $\epsilon_{3}^{e(h)}$ and $\epsilon_{8}^{e(h)}$ within the spectrum, cf. the black circles in the energy term schemes. In addition, all orbital energies of the $(2,8)$ configuration are enclosed in a narrower energy interval compared to $(3,7)$ whereas the energy spectra do not reveal any degeneracy.
However, for (2,8) the spectrum separates into two parts of similar energetic substructure with orbitals energies $\epsilon_0^{e(h)}$ to $\epsilon_4^{e(h)}$ and  $\epsilon_5^{e(h)}$ to $\epsilon_9^{e(h)}$, respectively.
Accompanying this fact, one clearly recognizes a change and an increase of the orbital symmetry when crossing over from the $(3,7)$ to the $(2,8)$ configuration. In contrast to $(a)$ the rotational and specular (mirror) symmetry of $\phi_{i,d^*,\lambda}^{e(h)}(\vec{r})$ with respect to perpendicular space axes in $(b)$ is increased. Moreover, the structure of the nodes (white lines with zero amplitude in figure ~\ref{fig.N10spectrum} right) of the HF orbitals changes, making the symmetry axes obvious. Particularly, inner and outer shell are clearly more separated by nodes in the $(2,8)$ configuration.

\section{Discussion and outlook}
In this paper we have considered ground state and dynamical properties of mesoscopic classical and quantum mass-symmetric electron-hole bilayers. In particular, we focused on the dependence of the properties on the layer separation $d$. The main effect is the gradual transition from systems with Coulomb interaction in the layers (at large $d$) to a system with short-range dipole interaction (at small $d$). Based on extensive classical molecular dynamics calculations we have shown that, with variation of $d$, several clusters show a sudden change of the ground state shell configuration, including several cases of re-entrant configuration changes which are related to symmetry properties. Furthermore, we have analyzed the classical normal modes of these bilayers and studied the $d$-dependence of the spectrum for $N=19$ as a representative example .

A striking result is the energy jump of the inter-shell rotation mode frequency $\omega_{SR^+}^2$ by more than four orders of magnitude when the ``magic'' ground state configuration (1,6,12) is replaced by (1,7,11).  
This leads us to suggest a new possibility for external control of inter-shell rotation by \textit{exerting strain} on the bilayer system (or alternatively by changing the trap frequency $\omega_0$ by an external electric field \cite{Xtrap}), i.e., a scheme which does not require changing particle number \cite{bonitz02,vova04}. Preparing a sample with $d$ slightly above $d_{cr}$, rapid compression initiates a ground state transition and thus allows to ``turn on'' the inter-shell rotation of composite dipoles --- excitons. Combined with optical excitation this may have interesting applications manipulating coherent emission.

In the second part of this paper we performed a quantum many-body calculation of the same system within the frame of a self-consistent Hartree-Fock approach. In the low-density limit  where the particles are well localized the classical properties are recovered. On the other hand, upon density increase and growing particle overlap quantum diffraction and exchange effects become important. This has significant consequences for the ground state phase diagram which is much richer than the classical one. There appear new structural phases which are characterized by charge localization on the outer shell coexisting with delocalization on the inner shell. Also, there exist parameter ranges where the classical and quantum systems show opposite shell configuration changes. The main advantage of the quantum many-body calculations is that they yield the complete single-particle energy spectrum and orbital-resolved ground states. We have shown that, even in the Wigner crystal phase where the density shows strong peaks, single peaks do not one-to-one correspond to single particles. On the contrary, in general, several orbitals contribute to a single density peak.

We note that the present quantum results correspond only to the lowest level of many-body theory --- the Hartree-Fock approximation. Thereby all pair interactions have been self-consistently included and direct and exchange terms are treated on the same footing. We have performed several comparisons with first-principle path integral Monte Carlo simulations which showed that the correct shell configurations are observed. This lets us expect that the quantum results reported in this paper will not change qualitatively when better approximations are being considered. Naturally, the first improvement to be made is the inclusion of scattering effects on the level of the second Born approximation of nonequilibrium Green's functions theory, as it was done e.g. in references \cite{bonitz96,kwong98,dahlen05}. We presently develop these calculations which will be reported elsewhere.

\ack
We wish to thank our colleagues S. Bauch and C. Henning for stimulating discussions. Part of this work was supported by the U.S. Department of Energy award DE-FG02-07ER54946 and the Innovationsfond Schleswig-Holstein.

\end{document}